# Assessment of Retracked Sea Levels from Sentinel-3A Synthetic Aperture Radar (SAR) Mode Altimetry over the Marginal Seas at Southeast Asia

Nurul Hazrina Idris[1,2], Stefano Vignudelli[3] and Xiaoli Deng[4]

[1]Tropical Resource Mapping Research Group, Department of Geoinformation, Faculty of Built Environment and Surveying, Universiti Teknologi Malaysia, 81310 Skudai, Johor Darul Takzim, Malaysia

[2]Geoscience and Digital Earth Centre, Research Institute for Sustainability and Environment, Universiti Teknologi Malaysia, 81310 Skudai, Johor Darul Takzim, Malaysia

[3] Consiglio Nazionale delle Ricerche (CNR), Area delle Ricerca CNR S.Cataldo, Via Moruzzi 1-56100 Pisa, Italy

[4]School of Engineering, The University of Newcastle, New South Wales, Australia

Corresponding Email: nurulhazrina@utm.my

This paper presents the assessment of altimetric data from Sentinel-3A satellite operating in Synthetic Aperture Radar (SAR) mode for sea level research studies and applications over the largest archipelagos at Southeast Asia. Both qualitative and quantitative assessments are conducted by analysing the physical shapes of waveforms, comparing with quasi-independent geoidal height data and independent tide gauge measurements. The results identified the percentage of ocean like and non-ocean like waveforms are 91% and 9%, respectively. Off 9% of non-ocean like waveforms, the major class is multi-peak (7%) followed by the quasi-specular waveforms (2%) observed near the coastline (<10 km). Ocean like waveforms typically appear beyond 500 m from the coastline. When comparing with geoidal heights and tide gauge measurements, the performance of sea levels from several retrackers are assessed. The SAMOSA+ retracker outperforms other retrackers (i.e. sub-waveform and modified threshold retrackers with 30%, 20% and 10%). That is, the standard deviation of differences against geoidal heights, and the temporal correlation against tide gauges are superior in most cases. In terms of root mean square error (RMSE), all retrackers are ranging with RMSE

≤20 cm in all cases. It can be concluded that in general, the SAMOSA+ retracked sea levels are accurate over the complicated regions at the Southeast Asia.

Keywords: SAR Altimetry, Coastal Waveform Retracking, Sea Level, South China Sea

**1.0 Introduction**

Synthetic Aperture Radar (SAR) altimetry over the ocean has attracted considerable attention since 2010 and remarkable progress has been made over the past years. Advances in data processing, combined with technological progress such as the advent of SAR altimetry from Cryosat-2 and Sentinel-3A/3B, have yielded more accurate retrievals of the geophysical parameters (i.e. sea level anomaly, SLA; significant wave height, SWH; and wind speed) in coastal zones, typically within several hundred meters from the coastline.

The advanced techniques of delayed Doppler SAR altimetry technology improves the estimation of geophysical parameters particularly over coastal oceans and inland water (Ray et al. 2013). Differing from the low resolution mode (LRM) conventional altimetry (e.g. Jason series and Envisat) that is based on the pulse limited footprint, in which the returned signal dictated by the length of the pulse, the SAR altimetry offers two independent variables in along and across satellite tracks. In the along-track direction, each resolution cell can be exploited using the phase information of the complex echoes that returned from the surface to satellite. The signal is characterised by a Doppler frequency. Therefore, an independent look can be referred to reduced footprint, which is effectively beam limited (Raney 1998, Ray et al. 2013).

In the past several years, the improvement in the estimation of geophysical parameters using SAR altimetry has been reported by many researchers (e.g. Jain et al. 2014, Ray et al. 2013, Ribana et al. 2017). Along the satellite track, no-data gap can be reduced until several hundred meters from the coastline, compared to ~7-10 km with the LRM altimetry (Andersen and Scharroo 2011, Bao et al. 2009, Deng and Featherstone 2002, Idris, Xiaoli, et al. 2017). The improved accuracy is obtained through the development of new and improved SAR altimetry retracking algorithms in several research and development projects such as the SAMOSA (SAR Altimetry Mode Studies and Applications) and SCOOP (SAR Altimetry Coastal and Open Ocean Performance) funded by the European Space Agency (ESA).

Similar to LRM altimetry, the requirement of specialised retrackers for SAR waveforms is vital in improving the estimated ocean parameters. The waveform retracking is a post-processing protocol to convert waveforms into scientific parameters of power amplitude (related to wind speed), range (related to sea level), and slope of leading edge (related to SWH) that characterise the observed scene (Gommenginger et al. 2011).

However, several issues particular to SAR altimetry remain open, specifically on the land contamination effects when the satellite track is parallel to the shoreline. Close to the coastline, SAR altimeter simultaneously views scattering surfaces of both water and land, producing complicated waveform patterns. The geometry of coastline, relief of land, and nature of terrain all contribute to defining the shape of waveforms. Although the SAR altimetry technology offers a high spatial resolution (300 m) at the along track direction, across the satellite track, the spatial resolution (~7 km) is still similar to those of the altimetry LRM waveforms (Dinardo et al. 2013, Thibaut et al. 2014). Due to extreme diversity of costal topographic characteristics throughout the world, a huge range of waveform shapes is thus observed in coastal areas.

These complex waveforms pose a real challenge to today's approach to retrack waveform. In the ESA funded project of CryoSat Plus for Ocean (CP4O), extensive study has been carried out to investigate different aspects of the opportunity offered by the new technology of SAR altimetry. This includes the improvement of the state-of-the-art SAMOSA SAR retracker that discards the delay/Doppler bins affected by land (Dinardo et al. 2018). The reduced bin of waveforms is subsequently retracked by SAMOSA retracker to obtain the geophysical parameters. The retracker is hereafter called the "SAMOSA+". The idea of discarding the bins affected by land has been previously applied to the LRM altimetry waveforms by Idris and Deng (2012), Passaro et al. (2014), and Bao et al. (2009), and also used on the Red3 retracker (Mercier F et al. 2010). It is proven as a robust technique when compared to the standard LRM full-waveform retracker (e.g. Maximum Likelihood Estimator 4, MLE4). However, taken into account the diversity of waveform patterns over coastal zones, the protocol has to be designed robust enough to exclude the extreme influence of land on the waveforms. This issue remains challenging to the field of both SAR and LRM altimeters.

Similar idea is proposed by García et al. (2018), but being applied to SARIn mode (applicable to CryoSat-2 only) when reconstructing the contaminated waveforms. Another approach is reducing the along-track resolution down to the theoretical limit equal to half the antenna length via fully focused altimetry processing (Egido and Smith 2017).

When developing a retracker for SAR waveforms, there are additional issues being the retrieval of sea level and SWH from the waveform sensitive to the error in the mispointing angle and to the effects of swell and swell direction. Therefore, the mispointing value input to the SAR retracker must be as accurate as possible (Gommenginger et al. 2013). This issue is still challenging to the SAR altimetry technical community. The sensitivity of retrieved parameters to errors in mispointing remains to be fully investigated (Gommenginger et al. 2013).

In what follows, this study conducts a novel assessment to identify the performance of the SAR altimetry retracked SLA over the complex marginal seas at the Southeast Asia. While regional assessment of SAR mode altimetric data are available worldwide (e.g. Labroue et al. 2012, Bonnefond et al. 2018, Peng and Deng 2020, Dinardo et al. 2018), only limited studies are available in the marginal seas at Southeast Asia (e.g. Passaro et al. 2016). This region is an important test bed on the Asian side where there are many islands and intricate coastlines. The typical waveform shapes are identified, and the performance of several retrackers are evaluated in this study. The used retrackers are SAMOSA+, sub-waveform, and modified threshold with 30%, 20% and 10% hereafter called "ModTh30", "Modth20" and "Modth10".

This paper is organized as follows: study area is described in Section 2; data sources and methods are provided in Sections 3 and 4, respectively; results and analysis is reported in Section 5; and discussions and conclusions are given in Section 6.

**2.0 Study Area**

The study area is situated at the marginal seas at Southeast Asia (Figure 1). It is one of the largest archipelagos in the Western Pacific oceans consisting of various coastal topographies such as beaches, mangroves, small islands and peninsulas. The Southeast Asia is surrounded by semi-closed ocean including the South China Sea, Strait of Malacca, Andaman Sea, Banda Sea, Natuna Sea, and Sulu Sea.

The area comprises a large range of ocean depth. The deep portion, called the China Sea Basin, has a maximum depth of 5,016 m and an abyssal plain with a mean depth of about 4,300 m. The climate is dominated by the seasonal tropical monsoon. In summer, monsoonal winds blow predominantly from the southwest; in winter, winds blow from the northeast. Annual rainfall varies from about 80 inches (2,000 mm) to more than 120 inches (3,000 mm) around the southern basin. Summer typhoons are frequent (Zhang et al. 2004).

## 3.0 Data Sources

Satellite altimetric data from Sentinel-3A SAR mode are obtained from the ESA Grid Processing on Demand (G-POD). The G-POD Service of SARvatore (SAR Versatile Altimetric Toolkit for Ocean Research & Exploitation) is the platform used to download L1B data products, which consist of multi-looked SAR mode waveforms with 512 gates. The high (20 Hz) posting rate data are obtained from July 2016 to April 2018 (Cycles 7 to 30). The SAMOSA+ retracked datasets, geophysical and environmental corrections are extracted for producing SLAs.

For the purpose of comparison, the quasi-independent geoidal height from the European Improved Gravity Model of the Earth by New Techniques 6C4 model (EIGEN6C4) are used (Förste et al. 2014). The geoidal height model is a static global combined gravity field model up to degree and order 2190 (1/8 degree grid), created from a combination of a multitude of data. They are available in the L1B data products.

Data from eight (8) tide gauge stations (Table 1) are also utilised for validation purpose. The hourly data are obtained from the University of Hawaii Sea Level Centre (https://uhslc.soest.hawaii.edu/). Table 1 summarizes the location of tide gauges, the Sentinel-3A satellite passes close to the tide gauge stations and their minimum distance.

## 4.0 Methods

The methodology of the study includes: 1) waveform classification using K-Mean clustering technique to understand the diversity of waveform classes on the experimental regions; 2) waveform retracking using reduced gate number of waveforms from sub-waveform (Idris 2014, Idris and Deng 2011, Idris et al. 2012), and modified threshold (i.e. Modth30, Modth20 and Modth10) (Lee et al. 2008) retrackers; and 3) retrieving sea levels from the retracked data, and tide gauges.

Note that when conducting the analysis based on the distance to coastline, the distance is computed along the satellite tracks.

## 4.1 Waveform Classification Using K-Mean Clustering

The waveforms are classified using K-mean clustering. It is an unsupervised classification technique partitioning the observations into several clusters. The clustering is performed by minimizing the sum of squared distances (e.g. Euclidean distance) between the observations and the corresponding centre of clusters (Tou and Gonzalez 1974).

In practice, the K-mean clustering requires the number of clusters to be specified. Since the number of waveform classes in the experimental region is unknown, the number of cluster has been varied from 6 to 15. When the number of cluster is 6, only limited number of classes is discovered. Initially, applying the K-mean clustering with 15 clusters is found appropriate for the datasets. That is, the clusters are better separated.

However, it is seen that the clustering algorithm produces several classes with very similar shapes. For instance, it produces several classes of the ocean + peaky trailing edge with varying land-like peak locations. Therefore, similar classes are grouped, resulting to 3 generic classes in the experimental regions. Detailed results are shown in Section 5.1.

## 4.2 Waveform Retrackers

The accuracy of SLAs from several retrackers are examined. They are the freely available SAMOSA+ retracked SLAs from the G-POD services, and the sub-waveforms, Modth30, Modth20 and Modth10 retracked SLAs that are processed/retrieved in this study. While the SAMOSA+ retracker is specialised for SAR waveform, the sub-waveform, Modth30, Modth20 and Modth10 are the retrackers applicable to the LRM altimetric such as Jason series and Envisat. In this study, those LRM altimetric retrackers are adapted to the SAR waveforms.

It is noted that the SAMOSA+ retracker is applied to the single-looked waveforms, meanwhile the other four retrackers are applied to the multi-looked waveforms. The single-looked is referred to the different Doppler beams, also known as looks that are synthesised to focus on a given surface on the ground, while the multi-looked is the averaging of all these Doppler power beams (or single-looked) (Garcia et al. 2014, Raney 1998, Ray et al. 2013). The production of multi-looked waveform involves the procedure of "multi-looking" that simply finds the incoherent summation in along-track direction of the weighted or non-weighted square-law detected range-compressed Doppler beam (looks). For detailed information about the technical processing, refers to Dinardo (2013).

4.2.1 SAMOSA+ Retracker

The SAMOSA+ retracked SLAs are available from the G-POD Service of SARvatore. They are retrieved by retracking the single Doppler beam (or single-look power waveform) to a SAR waveform model.

The returned SAR waveform can be demonstrated by a physically-based model of "SAMOSA" (Ray et al. 2013). It offers a complete description of a SAR altimeter returned waveform from ocean surface, expressed in the form of maps of reflected power in delay-Doppler space (also known as stack) or expressed as multi-looked echoes (Ray et al. 2013). The SAMOSA retracker is capable of accounting for an elliptical antenna pattern, mispointing errors in roll and yaw, surface scattering pattern, non-linear ocean wave statistics and spherical Earth surface effects (Gommenginger et al. 2013, Halimi et al. 2014, Ray et al. 2013). It estimates the parameter of epoch (related to sea level), composite sigma (related to SWH), amplitude (related to wind speed) and mispointing angle. The SAMOSA retracker has been widely applied to SAR altimetry data on-board of Sentinel-3A and 3B for operational usage, thus being considered as a standard SAR retracker.

The SAMOSA+ retracker is an improved SAMOSA model. It is specialised for coastal waveforms, which are usually contaminated by land or rough coastal sea states. The SAMOSA+ retracker applies a dual step retracking. In the first step, the SWH is estimated using the standard SAMOSA model; in the second step, the SWH was set to zero and the third free parameter in the retracking comes the mean square slope, producing the retracked range and amplitude (Dinardo et al. 2018).

4.2.2 Sub-waveform Retracker

Previous studies using LRM altimeter found excellent performance of sub-waveform retracker over several coastal regions including the Great Barrier Reef in Australia (Idris 2014), the Prince William Sound in Alaska the United States (Idris, Deng, and Idris 2017), and the marginal seas at Southeast Asia (Idris 2020). This retracker is found capable of reducing no-data gap near coastal/islands and improving the accuracy of estimated SLAs when compared to the standard MLE4 retracker (Thibaut et al. 2010).

In what follows, the sub-waveform retracker is adapted to the SAR waveforms. The procedures are: 1) extracting the truncated waveform, which concentrates on the ocean-like signal (and exclude the non-ocean like signal from the waveform). The waveform's leading edge and possible peaks are determined by waveform differential technique from Lee et al. (2008); 2) optimizing the parameters by concentrating on the truncated part of ocean-like waveforms without attempting to explain the full shape of waveforms using the simplified SAR model (Garcia et al. 2014), that is simple, but approximate, analytic model. Using the model, multi-looked waveforms are fitted to estimate the three parameters (i.e. epoch, composite sigma and amplitude).

Unlike other sub-waveform retracker such as Red3 (Mercier F et al. 2010) and ALES (Passaro et al. 2014), our sub-waveform retracker (Idris et al. 2012) takes into account the fact that the sub-waveform gates vary depending on the existence of the peaks on the trailing edge. The technique is found particularly robust to consider the varying locations of the land-like peak on the trailing edge.

However, one might concern about the accuracy of the optimized geophysical parameters when the number of gates are reduced. Therefore, further assessment is conducted to identify the effects of reducing the waveform gates. First, by assuming that the SAMOSA+ retracked SLAs are accurate over the open oceans, about 50,000 waveforms are gathered there. Second, these waveforms are retracked using the simplified SAR retracker (Garcia et al. 2014) by reducing the number of gates gradually to 360. Note that the gate number of 512 is the full waveforms, where all gates are considered during retracking. Third, the sub-waveform retracked SLAs are computed. Finally, the root mean square error (RMSE) are computed against the standard SAMOSA+ retracked SLAs for each reduced gate numbers (Figure 2).

From Figure 2, the value of RMSE is found large (31 cm) when the number of gates (360) in the trailing edge is totally discarded. The RMSE slowly decays to ≤7 cm when reduced gate is 380, and significantly unchanged or saturated until the reduced gates is 512. Based on the results, it can be inferred that the accuracy of the retracked SLAs using full-waveform (512 gates) is insignificantly different to those of the reduced gate 380. When reduced gates are smaller than 380, the sub-waveform retracker should be used in caution.

Examples of sub-waveform retracking with varying SWH are shown in Figure 3. It should be considered that the misfit between the power of sub-waveform retracking model and the power of waveforms data are reasonably high exceeding 0.9900.

### 4.2.3 Modified Threshold Retracker

Modified threshold retracker applies similar principles to those of sub-waveform retracker, in which the threshold algorithms (Davis 1997) are applied on waveforms reduced gate number (Lee et al. 2008). The procedures are: 1) the waveform's leading edge and possible peaks are determined by waveform differential technique from Lee et al. (2008); and 2) the mid-point of leading edge associated to sea levels is identified by applying the threshold retracker (Davis 1997), with threshold value of 10% (Modth10), 20% (Modth20) and 30% (Modth30).

### 4.3 Retrieval of Retracked Sea Levels

When performing waveform retracking, the main parameter of interest is the epoch ($t_0$). The retracked epoch ($t_0$) must be referred to the nominal tracking gate ($g_0$), which can be identified as follows (Dinardo, Personal Communication 2019):

$$t_i = -dt(N_{start}/2): dt: dt\,(N_{last} - 1)/2 \qquad (1)$$

$$dt = 1/(B \times O_{sampling}) \qquad (2)$$

$$I_i = t_i(1/dt) \qquad (3)$$

$$N_{start} = N_{FFT} \times O_{sampling} \qquad (4)$$

$$N_{last} = N_{FFT} \times O_{sampling} + E_s \times O_{sampling} \times 2 \qquad (5)$$

where $t_i$ is an array with step $dt$ with $i=1,2,3…512$, $B$ is bandwidth (320e6 Hz), $N_{FFT}$ is number of samples (128), $dt$ is the sampling time step and $O_{sampling}$ is over sampling. If zero padding is selected, $O_{sampling} = 2$; otherwise, $O_{sampling} = 1$. If extended window is selected, extra sample, $E_s=128$; otherwise, $E_s=0$. The nominal tracking gate ($g_0$) is where $I_i=0$. The retracked range ($R_{retracked}$) from sub-waveform and modified threshold is extracted using Equation 6 (Dinardo et al. 2018):

$$R_{retracked} = \left(R + (t_0 - g_0)\right) c/2 \qquad (6)$$

where $c$ is the speed of light, and $R$ is the on-board tracker range that has all the instrumental corrections already applied (e.g., Doppler range shift, ultra-stable oscillator and centre of gravity). The $R_{retracked}$ from SAMOSA+ retracker is obtained directly from the L1B data product, without implementing Equations 6.

The retracked SLAs can be obtained by subtracting the orbital height ($H$), retracked_range ($R_{retracked}$), corrections and mean sea surface ($mss$, Equation 7) (AVISO 2012):

$$SLA_{retracked} = H - R_{retracked} - \sum_{i=1}^{n} correction_i - mss \qquad (7)$$

The corrections are the dry and wet tropospheric from the European Centre for Medium-Range Weather Forecasts (ECMWF) numerical prediction model, inverse barometric and dynamic atmospheric correction. The ionospheric correction from General Ionospheric Model (GIM) is used instead of the instrumental-based dual-frequency ionospheric correction is due to the possible errors related with the land contamination in the Ku and C bands (Fernandes et al. 2014). Although dual frequency ionospheric correction is reported as the best solution for a global scale (Yang et al.

2019), it is more convenient to use a model-based correction for our case where the coastal topography is highly complicated.

The ocean tide is corrected using a harmonic analysis method (Pawlowicz et al. 2002) called as the "T-Tide". It modelled the tidal characteristics from the altimetric time series as the sum of a finite set of sinusoids at specific frequencies related to astronomical parameters. This alternative method is used instead of the state-of-the-art ocean tide model such as FES2012 and GOT4.8 to better model the tidal signals over coastal oceans (Idris et al. 2014). The mean sea surface is from DTU15 model. The 1 Hz geophysical corrections are interpolating at 20 Hz times. The sea state bias correction is not applied because it is not appropriate for waveforms near the coasts (Andersen et al. 2011).

When comparing the retracked SLAs against tide gauge, the hourly tide gauge data are interpolated onto altimeter time to ensure time gaps are minimised. Their mean differences of time series are removed for consistency. To remove the high frequency signals, both SLAs from altimeter and tide gauge are subtracted from the dynamic atmospheric correction. For correcting tidal effects from the SLAs, harmonic analysis method (Pawlowicz et al. 2002) is applied to both tide gauge and altimetric SLAs time series. The SLA time series are decomposed into its tidal components (tidal harmonic constituents) and then removed from the time series. Note that the global tide model corrections (e.g. FES 2014) available from the data product is not applied because previous studies (Andersen et al. 2011, Idris et al. 2014) highlighted the issue with this global model.

**5.0 Results and Analysis**

This section reports the analysis of SAR mode waveforms (Section 5.1), comparison of retracked SSHs with geoidal heights (Section 5.2) and comparison of retracked SLAs with tide gauges (Section 5.3).

## 5.1 Analysis of SAR Mode Waveforms

Figure 4 presents example of SAR mode waveforms along pass 303 (Figure 4a) near east coast of Peninsular Malaysia. The waveform echogram (Figure 4b) shows the shape of waveform became noisy when approaching the coastline. The standard ocean SAR waveforms are observed at 7.3 km from the coastline, however, SAR mode waveforms with peaks are observed at ~2 km from the coastline (Figure 4c). At location (2) in Figure 4a (corresponding to blue line in Figure 4c), the across-track distance to coastline is 2.1 km and along-track distance is 5.3 km. At this location, land impact is clearly seen on the waveforms. Although accurate altimetric measurements are expected typically within several hundred meters from the coastline, it is not the case when the satellite track runs parallel to the shoreline. The accuracy of measurement degrades particularly within ~7 km from coastline because of the across-track spatial resolution is sparse (~7 km), similar to those of the LRM satellite altimeters, leading to erroneous estimation in the altimetric signals.

Assessment based on waveform types provided in L1B data product indicates that ocean-like waveforms are available typically beyond ~500 m from the coastline. The percentages of ocean-like waveform and non-ocean like waveforms (Table 2) are ~91% and ~9%, respectively. Figure 5 shows the distributions of ocean-like (in light grey) and non-ocean like waveforms (in red). Non-ocean like waveforms are mainly distributed near coastlines. They also are found at complicated oceans topography around the Philippines and Indonesia, where many small islands exist. Their coastlines are highly dynamics with many physical processes such as Indonesian Throughflow, equatorial origin Rossby and Kelvin waves and seasonal upwelling. Note that the non-ocean waveforms are also found on the land showing the waveforms over inland water.

Detailed analysis of waveform peakiness and signal to noise ratio (SNR) in each 2 km bands from the coastline are provided in Figures 6a and 6b, respectively. The SNR provide information about the level of noise in the waveforms, which is computed by counting the ratio of mean and standard deviation. The higher the SNR, the more information is available. When approaching coastlines, waveform peakiness increases, meanwhile SNR decreases. The values of peakiness within 4-10 km bands are remain low with ~0.04 but suddenly spike up to ~0.1 within <2 km band. In contrast, the value of SNR within 4-10 km bands is ~0.45 but spike down to ~0.3 within 2 km band. These suggest the waveform's level of noise increases significantly within 2 km band.

Waveform classification using K-mean clustering technique indicates three generic classes (Figure 7): ocean like waveform (Class 1); multi-peak waveform (Class 2); and quasi-specular waveform (Class 3). Based on results in Figure 6c, percentage of Class 1 exceeds 95% beyond 2 km from coastline. However, in the last 2 km band, the percentage of Class 1 reduces to 73%. There, Classes 2 and 3 are observed with percentage of 25% and 2%, respectively. Class 2 is also observed up to 10 km bands, while Class 3 is only observed within 2 km band. Class 3 is usually observed over calm water such as over estuarine. These suggest that precise and accurate geophysical parameters such as the SLA and SWH can be observed beyond 2 km from the coastline. Technically, waveform Class 2, which consists of a clear leading edge, can be retracked effectively using sub-waveform retracker (such as SAMOSA+) that excludes the waveform bins contaminated by lands during the retracking process.

Previous study by Abdullah et al. (2017) reports an excellent altimetric data coverage (>85%) over the South China Sea, Gulf of Thailand and Straits of Malacca. However, low percentages are reported over Andaman Sea and Sulu Sea where data availability is between 64-76%. Study by Lumban Gaol et al. (2018) reported that the percentage of altimetric valid data from Jason-2 X-TRACK products are only 34% over continental shelf around Indonesia, where Brown and peaky waveforms are usually observed. This result is supported by Sinurat et al. (2019), in which over Halmahera Sea, Indonesia, the percentage of non-ocean like waveforms is 63% in shallow and narrow bay waters, and 10% around small island waters.

Recent study by Idris (2020) reports that the standard MLE4 retracker from the Sensor Geophysical Data Product (SGDR) of Jason-2 satellite only provides <20% of data around the Southeast Asia coastlines. In contrast, a systematic retracking expert system based on Fuzzy logic, the "CAWRES" (Idris, Deng, Md Din, et al. 2017, Idris and Deng 2013), can improve the percentage of reliable data to more than 80%. Note that comparison with other studies from SAR mode altimetric is impossible because there are no such studies have been carried out in the region.

## 5.2 Comparison of Retracked SSHs with Geoidal Heights

In this section, the precision of sea levels from different retrackers are compared to geoidal heights. To be consistent with geoidal height, parameter of sea surface height (SSH) is used instead of SLA. Unlike the SLA that refers to mean sea surface, both the geoidal height and SSH are based on World Geodetic System (WGS84) referenced ellipsoid. In this study, the mean and standard deviation of difference (STDs) are computed among both datasets. They are computed in each 2 km bands for up to 10 km from the coastline.

Figure 8 shows the comparison of the SAMOSA+ retracked SSHs with geoidal heights. Throughout the distance bands, the mean residuals are less than 150 cm. When getting closer to coastline, the STDs increase from 50 cm to 270 cm. This indicates reduction in terms of data precision.

It seems that the values of STDs computed in the Southeast Asia is somewhat larger than in other regions, for instance the Great Barrier Reef in Australia, where the STDs are between 14-91 cm (Idris, Deng, Md Din, et al. 2017), and the Prince William Sound in Alaska the United States, where the STD values are 11-69 cm (Idris, Deng, and Idris 2017).

Table 3 shows example of STDs computed for several retrackers over selective satellite passes. They are computed from SSHs close to coastlines (up to 10 km), and averaged within 2 km distance bands.

Based on results in Table 3, in general, SAMOSA+ retracked SSHs can provide precise data close to coastline. For distance band <2 km, off five cases, STDs of SAMOSA+ retracker are superior to those of other retrackers. Similar performance is also observed when distance bands are beyond 2 km from coastline, except for a few cases when sub-waveform and Modth30 are slightly better. Over satellite passes 231 and 117, STDs of SAMOSA+ retracker outperform other retrackers in all distance bands.

Figure 9 shows examples of SSH profiles along passes 117 and 303. Along pass 117 (Figure 9a), noisy SSHs are observed between longitudes 117.75º and 117.8º when the satellite passing through coral atolls. Along pass 303 (Figure 9b), the SSH profile fluctuates between longitudes 104.1º and 104.14º showing imprecise SSHs when passing near islands. Results in Figure 9 prove that the precision of altimetric SSHs is not only influenced by the coastal topographies, but also to the sea bottom features.

## 5.3 Comparison of Retracked SLAs with Tide Gauges

The retracked SLAs are validated against tide gauge data by computing the correlation coefficient ($r$) and RMSE. The values of $r$ and RMSE are averaged in each 2 km bands, up to 10 km from coastlines. Table 4 summarises the validation results that are computed for several retrackers.

Based on the results, small correlations are observed from all retrackers in four out of eight cases. The *r* values are <0.5 at Manila, Sabang, Waikelo and Davao stations suggesting that the retracked SLAs explain on average <50% of tide gauge total variance. Although the correlation coefficient is small, the value of RMSE is reasonably acceptable (<15 cm), suggesting the error of retracked SLAs are minor. The reason that might be responsible to the small correlation is because those three stations are located at complicated coastal topography with many small islands. In addition to the severe impact of land that results in inaccurate altimetric measurements, the geophysical corrections (e.g., geocentric tides and atmospheric models) are also inaccurate over these regions. Issue regarding the accuracy of geophysical parameters in deriving SLAs has been discussed in detailed by Andersen et al. (2011).

Comparison with Ambon, Ko Lak and Ko Taphao Noi and Bitung stations found that in general, SAMOSA+ retracker is superior to those of sub-waveform, Modth30, Modth20 and Modth10 retrackers. It records the highest *r* value in most cases. However, the RMSE values are insignificantly different than others. Only small differences are observed among the RMSE of those retrackers. In all cases, RMSE values from all retrackers are <20 cm indicating that they are reliable.

It is observed that the performance of empirical-based retrackers (e.g., Modth10) is always inferior to those of physical retrackers (i.e. SAMOSA+ and sub-waveform). For example near Ambon station, the correlation coefficient of Modth10 retracker is between 0.62-0.65, while SAMOSA+ and sub-waveform retrackers can provide *r* value between 0.71-0.73 and 0.70-0.78, respectively. This indicates that on average, the Modth10 describe only <0.65% of tide gauge total variance while SAMOSA+ and sub-waveform explain >70%.

Figure 10 shows the time series of SLAs for all retrackers along pass 053 near Ko Lak. The SLA profiles from tide gauge is also shown. It seems that the retracked SLAs have large variations, while the profile from tide gauge is much smoother. The correlation value is the highest from SAMOSA+ retracker at 8-10 km bands and reduces to 0.59 when approaching coastline.

## 6.0 Discussions and Conclusions

This paper analyses the performance of SAR mode altimetric data from Sentinel-3A in the largest archipelagos in the Western Pacific, where complicated coastal topographies are expected to influence the accuracy of the geophysical parameters. The performance of SAMOSA+, sub-waveform, and modified threshold with 30%, 20% and 10% is examined there. The assessments include the analysis of waveform, comparison with geoidal heights, and validation against tide gauge data.

The analysis on waveform patterns shows about 9% of waveforms are following the non-ocean like shapes, which can be further classified into multi-peak (with land impact on waveform's trailing edge) and quasi-specular waveforms. The values of peakiness (SNR) increases (decreases) when approaching coastlines. Given that the shape of waveforms depends on the track orientation, the angle of intersection with the coast, and the land topography (Aldarias et al. 2020), accurate geophysical parameters can be expected beyond 2 km from the coastlines, where ocean like waveforms typically appear.

When comparing the SSHs from SAMOSA+ retracker against geoidal heights and tide gauges, SAMOSA+ retracker can provide precise and accurate data close to coastline. In most cases, STDs and $r$ of SAMOSA+ retracker are superior to those of other retrackers. However, the RMSE indicates small differences ($\leq 5$ cm) among all retrackers indicating that all retrackers are reliable. It is also exhibited that the precision of SSHs is influenced by both coastal topographies and sea bottom features such as coral atolls.

High rate of waveform's contaminations in the similar study area has been reported by previous studies using LRM altimetric data such as Jason-2 (Idris 2020, Lumban Gaol et al. 2018, Sinurat et al. 2019) and SARAL/AltiKa (Abdullah et al. 2016, Mohammed et al. 2016). However, limited studies are reported the analysis using SAR mode altimetric there.

In identifying the performance of waveform retracker for the region, previous studies (Abdullah et al. 2016, Idris 2020, Mohammed et al. 2016, Sinurat et al. 2019) found that the empirical-based retracker (i.e. Ice or threshold) is superior to that of the standard MLE4 retracker (which is a physical-based retracker). This is in contrast to the finding from this study. Here, the physical-based retracker of SAMOSA+ is found superior to that of the empirical-based retracker. Excellent performance of SAMOSA+ from L1B product is mainly due to the fact that it is applied to single-looked waveforms, rather than multi-looked waveforms in this study.

Study by Passaro et al. (2016) that inter-compared the SLAs from Envisat (reprocessed with the ALES retracker) and Cryosat-2 SAR over Indonesian Seas has identified that the SAR retracked SLAs can help lower the level of uncertainty in the computation of sea levels. The SAR altimetry provides better measurements than Envisat altimetry, even without any coastal-dedicated retacking SAR waveforms.

The main reason that SAMOSA+ retracker is excellent when compared to other retrackers is because the SAR mode processing is based on the accumulation of 256 looks over each Doppler band (Ray et al. 2013). This is the advantageous of SAMOSA+ retrackers, which is reprocessed on single-looked of accumulated waveforms, meanwhile the modified threshold and sub-waveform retrackers are processed on multi-looked waveforms.

It can be concluded that the SAMOSA+ retracker, which is an improved SAMOSA retracker specialised for coastal ocean is vital for coastal applications at Southeast Asia, where worst-case scenarios are commonly observed on the waveform signals. The availability of accurate SAR altimetry data over this complicated regions enable the confidence to extend the use of the products to any coastal applications such as coastal sea level rise studies.

Although the study identified that SAMOSA+ retracker is effective for the study region, further analysis should be carried out to identify the accuracy of the retracked SLAs by considering the varying waveform classes. This is crucial particularly when multiple retrackers are considered in attempting to optimize the accuracy of retracked SLAs over coastal oceans.In addition, further comparison with other altimetric data at cross-over points can provide understanding on the consistency of the retracked SLAs over different missions.

**Acknowledgments**

The research is supported by the Fundamental Research Grant Scheme (Vot 5F262, Reference Code:FRGS/1/2019/WAB05/UTM/02/1), Ministry of Education Malaysia and Universiti Teknologi Malaysia, and Australian Endeavour Fellowship. We would like to acknowledge the European Space Agency (ESA) GPOD data team for kindly providing Sentinel-3A data, and the University of Hawaii for providing tide gauge data. Special thanks to Dr. Dinardo Salvatore from ESA for his expert advice particularly on the Sentinel-3A data usage.

**Table**

Table 1. Tide gauge stations and nearby Sentinel-3A altimeter ground passes (as shown in Figure 1). The minimum distance among both datasets are also shown.

| Tide Gauge Stations | Sentinel-3A Pass Number | Minimum Distance (km) |
|---|---|---|
| Ambon | 302 | 35 |
| Bitung | 074 | 13 |
| Davao | 017 | 4 |
| Ko Lak | 054 | 9 |
| Ko Taphao Noi | 289 | 30 |
| Manila | 003 | 18 |
| Sabang | 118 | 11 |

| | Waikelo | 231 | 7 |
|---|---|---|---|

Table 2. The percentage of waveform classes computed over the study region

| Waveform class | Percentage (%) |
|---|---|
| Ocean-like | 91.24 |
| Non ocean-like | 8.76 |

Table 3. Standard deviation of difference (STD in cm) of various retracked SSHs with respect to EIGEN6C4 geoidal heights for several passes. STDs are averaged values within 2 km bands computed up to 10 km. The smallest STD value for each distance band is indicated in bold.

| Satellite Pass No. | Waveform Retrackers | Distance Bands | | | | |
|---|---|---|---|---|---|---|
| | | <2 km | 2-4 km | 4-6 km | 6-8 km | 8-10 km |
| 303 | SAMOSA+ | **99** | 118 | **70** | **65** | **56** |
| | Sub-waveform | 258 | 124 | **70** | 67 | 58 |
| | Modth10 | 542 | 130 | 112 | 74 | 60 |
| | Modth20 | 393 | 118 | 82 | 74 | 60 |
| | Modth30 | 335 | **112** | 78 | 74 | 59 |
| 246 | SAMOSA+ | **50** | 66 | **69** | 64 | 40 |
| | Sub-waveform | 353 | 81 | 77 | **63** | 40 |
| | Modth10 | 550 | 59 | 113 | 95 | 40 |
| | Modth20 | 448 | 60 | 75 | 63 | 40 |
| | Modth30 | 410 | **58** | 75 | 64 | **39** |
| 289 | SAMOSA+ | **106** | 124 | **95** | **91** | 51 |
| | Sub-waveform | 259 | 133 | **95** | **91** | 50 |
| | Modth10 | 622 | 121 | **95** | **91** | 46 |
| | Modth20 | 435 | 121 | 96 | 92 | **45** |
| | Modth30 | 306 | **120** | **95** | **91** | **45** |
| 231 | SAMOSA+ | **200** | **140** | **34** | 40 | **28** |
| | Sub-waveform | 814 | 335 | 350 | 42 | 32 |
| | Modth10 | 1136 | 351 | 405 | 43 | 57 |
| | Modth20 | 963 | 345 | 397 | 42 | 53 |
| | Modth30 | 890 | 341 | 397 | **41** | 49 |
| 117 | SAMOSA+ | **232** | **63** | **47** | **29** | **64** |
| | Sub-waveform | 627 | 308 | 50 | 30 | 66 |
| | Modth10 | 840 | 339 | 50 | 32 | 92 |
| | Modth20 | 685 | 336 | 49 | **31** | 89 |
| | Modth30 | 686 | 321 | 48 | **31** | 86 |

Table 4. Mean of correlation coefficient (*r*) and RMSE (in cm) of retracked SLAs with respect to tide gauge stations computed in each 2 km bands from the coastlines. The highest value of *r* for each distance band is indicated in bold.

| Station / Distance to coast | | <2 km | | 2-4 km | | 4-6 km | | 6-8 km | | 8-10 km | |
|---|---|---|---|---|---|---|---|---|---|---|---|
| | Retrackers | r | RMS | r | RMS | r | RMS | r | RMS | r | RMS |
| Ambon & Pass 302 | SAMOSA | 0.72 | 19 | **0.71** | 9 | **0.73** | 19 | **0.73** | 19 | **0.73** | 20 |
| | Sub-waveform | **0.78** | 21 | 0.70 | 18 | 0.71 | 20 | 0.70 | 19 | 0.70 | 20 |
| | Modth10 | 0.64 | 19 | 0.65 | 18 | 0.62 | 18 | 0.64 | 18 | 0.63 | 19 |
| | Modth20 | 0.67 | 19 | 0.68 | 18 | 0.67 | 20 | 0.67 | 19 | 0.67 | 20 |
| | Modth30 | 0.70 | 19 | 0.69 | 18 | 0.69 | 20 | 0.69 | 19 | 0.69 | 20 |
| Ko Lak & Pass 054 | SAMOSA | 0.59 | 13 | **0.64** | 18 | **0.60** | 17 | **0.64** | 18 | **0.65** | 18 |
| | Sub-waveform | **0.60** | 25 | 0.60 | 18 | **0.60** | 18 | 0.62 | 18 | 0.63 | 18 |
| | Modth10 | 0.48 | 10 | 0.54 | 17 | 0.54 | 18 | 0.56 | 16 | 0.57 | 17 |
| | Modth20 | 0.53 | 10 | 0.59 | 18 | 0.59 | 18 | 0.59 | 17 | 0.60 | 17 |
| | Modth30 | 0.53 | 10 | 0.60 | 19 | 0.59 | 19 | 0.61 | 17 | 0.62 | 17 |
| Ko Taphao Noi & Pass 289 | SAMOSA | **0.62** | 14 | **0.63** | 14 | **0.62** | 14 | **0.63** | 13 | **0.64** | 12 |
| | Sub-waveform | 0.60 | 15 | 0.60 | 14 | 0.60 | 14 | 0.60 | 13 | 0.62 | 12 |
| | Modth10 | 0.52 | 14 | 0.52 | 14 | 0.52 | 12 | 0.52 | 13 | 0.54 | 11 |
| | Modth20 | 0.57 | 14 | 0.56 | 14 | 0.56 | 13 | 0.56 | 12 | 0.57 | 11 |
| | Modth30 | 0.58 | 14 | 0.58 | 15 | 0.58 | 13 | 0.58 | 13 | 0.60 | 11 |
| Bitung & Pass 074 | SAMOSA | - | - | - | - | **0.88** | 15 | **0.89** | 15 | 0.88 | 15 |
| | Sub-waveform | - | - | - | - | 0.87 | 15 | 0.87 | 14 | 0.88 | 15 |
| | Modth10 | - | - | - | - | 0.83 | 14 | 0.83 | 13 | 0.84 | 13 |
| | Modth20 | - | - | - | - | 0.84 | 14 | 0.85 | 14 | 0.87 | 14 |
| | Modth30 | - | - | - | - | 0.86 | 14 | 0.87 | 14 | **0.89** | 14 |
| Manila & Pass 003 | SAMOSA | 0.34 | 11 | 0.34 | 11 | 0.35 | 11 | 0.33 | 10 | 0.33 | 11 |
| | Sub-waveform | 0.35 | 11 | 0.31 | 11 | 0.33 | 11 | 0.31 | 11 | 0.27 | 12 |
| | Modth10 | 0.30 | 12 | 0.27 | 11 | 0.24 | 12 | 0.26 | 12 | 0.21 | 11 |
| | Modth20 | 0.31 | 12 | 0.31 | 12 | 0.29 | 11 | 0.31 | 12 | 0.27 | 12 |
| | Modth30 | 0.33 | 11 | 0.32 | 11 | 0.31 | 11 | 0.31 | 12 | 0.30 | 11 |
| Sabang & Pass 118 | SAMOSA | 0.49 | 12 | 0.50 | 13 | 0.50 | 13 | 0.50 | 13 | 0.50 | 12 |
| | Sub-waveform | 0.46 | 12 | 0.47 | 13 | 0.49 | 13 | 0.49 | 13 | 0.47 | 14 |
| | Modth10 | 0.40 | 12 | 0.40 | 13 | 0.42 | 13 | 0.43 | 13 | 0.39 | 13 |
| | Modth20 | 0.45 | 13 | 0.45 | 12 | 0.46 | 13 | 0.46 | 13 | 0.44 | 14 |
| | Modth30 | 0.46 | 13 | 0.46 | 12 | 0.47 | 12 | 0.48 | 13 | 0.45 | 13 |
| Waikelo & Pass 231 | SAMOSA | 0.52 | 14 | 0.49 | 13 | 0.50 | 12 | 0.49 | 13 | 0.52 | 13 |
| | Sub-waveform | 0.47 | 14 | 0.46 | 12 | 0.47 | 12 | 0.46 | 13 | 0.46 | 13 |
| | Modth10 | 0.36 | 14 | 0.37 | 14 | 0.32 | 13 | 0.35 | 14 | 0.34 | 13 |
| | Modth20 | 0.41 | 14 | 0.42 | 13 | 0.40 | 13 | 0.42 | 13 | 0.40 | 13 |
| | Modth30 | 0.46 | 14 | 0.44 | 13 | 0.43 | 12 | 0.44 | 13 | 0.44 | 14 |
| Davao & Pass 017 | SAMOSA | 0.01 | 15 | 0.14 | 14 | 0.10 | 13 | 0.07 | 13 | 0.17 | 13 |
| | Sub-waveform | 0.15 | 12 | 0.06 | 13 | 0.11 | 13 | 0.09 | 13 | 0.13 | 14 |
| | Modth10 | 0.31 | 17 | 0.30 | 13 | 0.03 | 12 | 0.05 | 13 | 0.04 | 12 |
| | Modth20 | 0.14 | 15 | 0.22 | 13 | 0.03 | 12 | 0.01 | 12 | 0.12 | 14 |
| | Modth30 | 0.04 | 15 | 0.18 | 13 | 0.11 | 13 | 0.05 | 13 | 0.15 | 14 |

**Figure Captions**

Figure 1. Study area is situated at the marginal seas at Southeast Asia. The Sentinel-3A satellite passes are shown in red lines, and the location of tide gauge stations are indicated in yellow marks. The satellite passes used in validation are shown in green lines and text (depicted from Google Earth, 2020).

Figure 2. The RMSE of the sub-waveform retracked SLAs against the SAMOSA+ SLAs for each reduced gate numbers. The value of RMSE (7 cm) is saturated starting from reduced gates 380 (red line) to gate 512.

Figure 3. Examples of sub-waveform fitting results when the number of gates is reduced to 380 at a) SWH 0.5 m, b) SWH 2 m, and c) SWH 6 m.

Figure 4. (a) Sentinel-3A pass 303 near east coast of Peninsular Malaysia. (b) Waveform echogram along pass 303. White colour indicates the land/island. (c) Examples of waveform shapes marked as number 1 (in green), 2 (in cyan) and 3 (in magenta), respectively in (a).

Figure 5. Spatial plot of waveform generic classes along the Sentinel-3A satellite tracks. The light grey and red colours show the ocean-like and non-ocean like waveforms, respectively. Note that the red marks on the land showing the waveforms over inland water.

Figure 6. (a) Waveform peakiness, (b) waveform signal-to-noise ratio, and (c) percentage of waveform classes: Class 1 is ocean-like waveform; Class 2 is multi-peak waveform; and Class 3 is quasi-specular waveform.

Figure 7. The waveform classes from K-Mean Clustering technique. (a) Class 1 is the ocean like waveform, (b) Class 2 is multi-peak waveform, and (c) Class 3 is quasi-specular waveform.

Figure 8. Mean (in black) and standard deviation (in grey) of residual among SAMOSA+ retracked SLAs and EIGEN6C4 geoid heights within 2 km bands from the coastline.

Figure 9. Examples of sea surface height (SSH) profiles from different retrackers and geoidal heights from Cycle 012 (a) Pass 117 and (b) Pass 303. The left side is the location of coastlines.

Figure 10. Time series of SLAs from several retrackers at difference distance from the coastline. They are from pass 053 near Ko Lak tide gauge stations. The time series from Ko Lak tide gauge station is also shown.